\documentclass[preprint,nofootinbib,superscriptaddress,eqsecnum]{revtex4}

\newcommand{\GeV}{{\rm GeV}}
\newcommand{\MeV}{{\rm MeV}}
\newcommand{\TeV}{{\rm TeV}}

\newcommand{\cm}{{\rm cm}}

\newcommand{\MBH}{\mbox{\rm MBH}}

\newcommand{\beq}{\begin{equation}}
\newcommand{\eeq}{\end{equation}}
\newcommand{\beqa}{\begin{eqnarray}}
\newcommand{\eeqa}{\end{eqnarray}}

\newcommand{\lsim}{\mathrel{\rlap{\lower4pt\hbox{\hskip1pt$\sim$}}
    \raise1pt\hbox{$<$}}}         
\newcommand{\gsim}{\mathrel{\rlap{\lower4pt\hbox{\hskip1pt$\sim$}}
    \raise1pt\hbox{$>$}}}         

\begin{document}


\vspace*{.0cm}

\title{QCD effects on ``stable'' micro black holes at the LHC}

\author{Itzhak Goldman}\email{goldman@wise.tau.ac.il}
\affiliation{ Department of Exact Sciences, Afeka College of
Engineering \\
 Bnei Ephraim 218, Tel Aviv 69107, Israel}
\author{Yuval Grossman}\email{yg73@cornell.edu}
\affiliation{ Institute for High Energy Phenomenology\\ Newman Laboratory of Elementary Particle Physics\\ Cornell
University, Ithaca, NY 14853, USA}
\author{Shmuel Nussinov}\email{nussinov@post.tau.ac.il}
\affiliation{ Tel Aviv University, Sackler School Faculty of Sciences \\ Ramat Aviv, Tel Aviv 69978, Israel}
\affiliation{ Chapman University, Schmid College of Science \\ Orange, California 92866, USA\vspace*{10mm}}

\vspace{1cm}
\begin{abstract}
If Micro Black Holes (MBHs) can be produced at the LHC, they will
decay very fast. We study hypothetical MBHs that do not decay; in
particular, QCD effects on accretion by MBHs that are produced at
rest. We explain why accretion of a nucleon by such MBHs is associated
with pion emission. This pion emission results in a kick to the MBHs,
such that their velocities are large enough to escape the Earth. Our
study provides an extra assurance that MBHs which might be produced at
the LHC are not dangerous.

\end{abstract}

\maketitle

\section{Introduction}

Hopefully, the LHC will reveal new physics at the TeV scale. One interesting possibility is that compactified extra
dimensions will manifest via $O(\TeV)$ Kaluza-Klein (KK) recurrences of Standard Model (SM)
particles~\cite{ArkaniHamed:1998rs,Antoniadis:1998ig,Randall:1999ee}. Some of these scenarios unify SM gauge
interactions and gravity at the Planck scale of the theory, $M^*_{\rm Pl}= O(\TeV)$. This suggested that Micro Black
Holes (MBHs) can be produced at LHC, providing direct evidence for the connection of the new physics with
gravity~\cite{Giddings:2001bu,Dimopoulos:2001hw} (see, however, \cite{Meade:2007sz}).

The esoteric possibility that some of these MBHs are stable and eventually accrete the whole Earth has been raised. In
a recent detailed paper Giddings and Mangano (GM)~\cite{Giddings:2008gr} showed that even if, contrary to basic physics
laws, the MBHs keep accreting and do not decay, the catastrophic scenario is experimentally excluded.  The strategy
adopted by GM is extremely conservative and straightforward. To avoid any assumptions which is not directly backed up
by observation they do not discuss the early stages of accretion by the $O(\TeV)$ initial MBHs. They show that MBHs
produced by high energy cosmic rays stop in white dwarfs and neutron stars and destroy them. The longevity of these
compact stars then completely and convincingly excludes all these problematic scenarios.

In the following we complement the analysis of GM addressing the first few accretions by the initial small MBH of
Schwartzschield radius \beq R_{\rm SW} \sim L^*_{\rm Pl} = \frac{1}{M^*_{\rm Pl}} \sim 10^{-17} \,\cm. \eeq Using a
very mild assumption, that is, that QCD can be used, we conclude that the accretion cross section is smaller than the
naive expectation. Moreover, the accretion is accompanied by pion emission. This results in a recoil for the MBHs that
eject them from Earth, and with a large probability, from the solar system as well. Thanks to the model-independent
work of GM no further arguments for the peaceful nature of putative LHC-produced MBHs are needed. Yet our independent
discussion is interesting in its own right.

We start by recalling the argument for the instability of accretion MBHs. Pair-produced MBHs carrying conserved charges
can be stable. GM, however, emphasized that electrically charged MBHs produced by ultra high energy cosmic rays will
stop in the Earth and consume it. This direct argument against charged MBHs suggests that strong electric fields exist
near charged MBHs and discharge them via $e^+ e^-$ pair creation. Conceivably we can still stabilize the produced MBHs
by endowing them with some topological ``charge''. This stabilizes only the ground state of the MBH, but not the more
massive ones that can be produced via accretion.

The key feature of a black hole is its ability to accrete any particle and increase its mass and corresponding
Schwartzschield radius. Consider then the accretion of a low energy
photon 
\beq \label{eq1}
     \MBH+  \gamma \to \MBH'.
\eeq 
Energy-momentum conservation requires that such an absorption is allowed only if the MBH has an excited state with
mass \beq \label{eq1prime}
    M(\MBH') =M(\MBH) + \omega,
\eeq with $\omega$ the photon energy. Since $\omega$ can assume any value, we conclude that the final state, $\MBH'$,
has a width, $\Gamma$, exceeding the energy $\omega$. This implies that $\MBH'$ should decay.

Hawking radiation provides precisely such a width. A MBH of mass $M$ and radius $R= M/M_{\rm Pl}^2$ has temperature
$T\sim M_{\rm Pl}^2/M$ and effective level spacing of $T$. This then allows the reaction (\ref{eq1}) to go so long as
$\omega \sim T$. Indeed the required width of $\MBH'$ satisfies $\Gamma(\MBH') \sim T$,  the rate of emission of
individual quanta in Hawking's ``Black body radiation'', $dN/{dt} \sim \Gamma \sim T$. 

One can still arbitrarily postulate discrete levels of stable MBHs of increasing masses and Schwartzschield radii. We
denote these discrete states as $\MBH_n$. Accretion of a nucleon, $N$, at rest by a slow MBH can now happen via the $2
\to 2$ kinetically viable process \beq\label{eq2-2}
  \MBH_n+ N \to \MBH_m+ X,
\eeq with $\MBH_m$ heavier than the initial state, $\MBH_n$. The mass difference is less than the mass of the nucleon,
$m_N \sim \GeV$. This is in order to allow emitting the final particle $x$ which can be a photon or pion.

Production of particles of large mass $M$ at the LHC is suppressed by the quick decrease of parton distribution
functions.  GM conservative upper limit for the total number of MBHs produced at LHC is about $10^{10}$ ($10^8$) for
$M(\MBH)\sim 1\;$TeV ($4\;$TeV) of which at most a fraction of about $10^{-4}$ will be produced with velocities less
than the escape velocity from Earth, $v^E_{\rm esc} \approx 11\;{\rm Km}/{\rm sec}$ or from the solar system, $v^S_{\rm
esc} \approx 45\;{\rm Km}/{\rm sec}$.

The cross section for accreting a particle of velocity $\beta$ on a large black hole is the geometric cross section
enhanced by gravitational ``focusing'' of slow, $\beta \ll 1$, particles \beq\label{Large-BH} \sigma_{\rm ac}^{\rm geo}
\sim \pi \, {R_{\rm SW}^2\over {\beta}^2}, \eeq where $R_{\rm SW}$ is the Schwartzschield radius radios of the MBH. For high
energy MBH of size $L^*_{\rm Pl} \sim \TeV^{-1}$, we can estimate the cross section by using $\beta \sim 1$ in
eq.~(\ref{Large-BH}). This yields the result used in~\cite{Giddings:2008gr} \beq \label{g-cs} \sigma_{\rm ac} \sim \pi
(L^*_{\rm Pl})^2 \sim 10^{-33}\, {\rm cm}^2. \eeq For the very slow MBHs which are gravitationally bound to the Earth
with $\beta \lsim 3 \times 10^{-5}$, Eq.~(\ref{Large-BH}) would imply very large accretion cross section of about
$10^{-24}\,{\rm cm^2}$. As we explain below, this estimate is not applicable for accretion of a nucleon. 

In this work we discuss the effect of QCD on such non-physical MBHs that do not decay. In particular, we show that they
are unlikely to accrete a nucleon, but rather, they interact in a two-body process which involves emission of a pion. The
outgoing MBH has then a recoil kinetic energy that is large enough to escape the Earth, and likely also the solar system.  Furthermore,  due
to QCD effects, the cross section for this process is strongly suppressed. The reason is that the MBH has to interact
with a di-quark and not with a single quark.

\section{QCD Effects in the first collisions}

In all the estimates above we assumed that the accreted object is point-like. This is, however, not a good assumption
in the first, subnuclear stage of accretion. Then, the nucleon radius, $R_{\rm N}\sim \GeV^{-1}$, is about
$10^3$ times larger than the size of the quantum MBH, $R_{\rm SW}\sim \TeV^{-1}$.  Hence the probability to accrete the
whole nucleon is highly suppressed. Accretion of a nucleon by slow MBH can happen via the ``completely point-like
configuration'' where all three quarks are in a sphere of radius $R_{\rm SW}$ with tiny probability suppressed by
$(R_{\rm SW}/R_{N})^6$. Naively, the most likely process to occur is that when the MBH accretes a point-like quark. As
we explain below, however, a single quark accretion is not possible. Thus, the most likely accretion process 
is when the MBH interacts with a di-quark inside the nucleus.

Next we show that a MBH cannot accrete a single quark.  The point is that if a single quark were accreted, the strong
chromostatic field quickly generates, via the Schwinger mechanism, $q-\bar q$ pairs in the region immediately outside
$R_{\rm SW}$. 
One anti-quark then falls onto the MBH and color neutralizes it. The quark, on the other hand, combines with the
``spectator'' di-quark from the initial nucleon forming a color singlet state with a baryon number one. The lightest
object with this baryon number is a nucleon. Thus, the original nucleon is regenerated and there is no accretion.

The above argument is modified for the case of a di-quark accretion. In that case the quark out of the $q -\bar q$ pair created combines with the MBH to form a color singlet, and the anti-quark combines with the spectator quark forming a meson. The lightest such meson 
is a pion.  In that case the MBH gains energy and accretion is
possible. The accretion therefore proceeds via the following $2 \to 2$ process 
\beq 
\label{dom-cross} 
\MBH+N \to \MBH' + \pi. 
\eeq 
A crucial point about the above process is that the resulting MBH has kinetic energy. In order to estimate this energy we note that the spectator quark along with the anti-quark created by the
Schwinger mechanism have energy of order $\Lambda_{QCD}$, that is, about a third of $m_N$. Thus, the pion momenta are
$k = O(300) \;{\rm MeV}$.  Since we care only about slow MBH, the Earth frame is also the center of mass
frame, and thus $\MBH'$ has a momenta similar to that of the pion. The corresponding velocity of the outgoing MBH is
\beq 
\beta={p \over M} \sim {300\;\MeV \over 1\;\TeV} \sim 100\; {\rm Km/sec}\,. 
\eeq 
This velocity is larger than the
escape velocity from Earth and even from the solar system. This leads to our basic observation: the first accretion
results in a kick to the MBHs that, in the absence of any other interaction, would send them out of the solar system.

We conclude this section with a few remarks:
\begin{itemize}
\item In principle, we can have more particles in the final states. For such cases the MBH kinetic energy is
    somewhat suppressed. Yet, the probability for such events are highly suppressed. The reason is that the available  energy released is less than half the nucleon mass, and thus it is very unlikely that another
    pion or photon would be produced. 
    
\item The probability to produce a final state with no pion at the final state is suppressed by
    $(R_{\rm SW}/R_{N})^3 \sim 10^{-9}$, and thus not completely negligible. Yet, it takes hundreds of accretions
    for the MBH to reach the size of the nucleon, where the volume suppression is not important. The
    probability that all of them will be without emission of a pion is negligible. 
    
    \item The nucleon can be inside
    a nucleus. This, however, does not affect our conclusion. The reason is basically that the binding energy is
    much smaller compared to the energy of the pion, and can be neglected.
\end{itemize}

\section{Cross sections estimates}

In order to see if indeed such a non-physical MBH will escape the Earth and the solar system we need to study its
energy loss due to scattering off matter. For this we estimate the accretion cross section, $\sigma_{\rm ac}$, the
scattering cross section, $\sigma_{\rm sc}$, and the kinetic energy transfer in each process, $\Delta E_{\rm ac}$ and $\Delta
E_{\rm sc}$. Using these cross sections and energy transfers, we can check if the MBH will escape the
Earth and the solar system.

We start with the estimate of $\Delta E_{\rm ac}$. We consider the dominant $2\to 2$ accretion, Eq.~(\ref{dom-cross}).
The kinetic energy transfer in each accretion is 
\beq 
\label{gen-acc} 
\Delta E_{\rm ac} = E_k({\rm out})-E_k({\rm in}) \approx {k^2 \over 2M} + {\vec p\cdot \vec k\over M} -{m_N \beta^2 \over 2},
\eeq 
where $M$ is the mass of the MBH, $k$
is the pion momentum, $p$ is the initial momentum of the MBH and $\beta$ is the initial velocity of the nucleon. The
last negative correction reflects the energy loss due to the gravitational binding of the accreted mass. The $O(\GeV)$
energy accreted in each step slightly increases the MBH mass and for the first few hundred collisions it is negligible.
Note that for the MBH that was created with very small velocities, $p$ is very small in the first accretion  and the MBH can
only gain kinetic energy. Yet, in subsequent accretions it may lose energy, depending on the relative direction of $p$
and $k$.

We next estimate $\sigma_{\rm ac}$. Our starting point is the naive geometrical cross section that applies to fast
MHBs, Eq. (\ref{g-cs}). There are several factors affecting this  that we discuss
below. Despite the fact that the quarks inside the nucleon are quasi-relativistic, the $1/{\beta^2}$ enhancement
for the accretion of a point-like non-relativistic particle, Eq.~(\ref{Large-BH}), applies. The reason is twofold:
During the time required for the MBH to traverse the nucleon, each quark traverses it back and forth $O(1/\beta)$ times
and therefore has $1/\beta$ chance to collide with the MBH.  The second $1/\beta$ factor stems from the exothermic $2
\to 2$ accretion reaction. The flux and phase space factors in the cross section yields roughly $\beta_f/\beta$ with
$\beta_f$ the velocity of the outgoing light particle. For elastic scattering this factor is one and thus irrelevant.
Yet, in our case, the final pion is relativistic and $\beta_f\sim 1$. Thus  we have an extra factor of $1/\beta$. Altogether the kinematic enhancement
factor is 
\beq 
\label{kin-sup} 
f_{\rm kin}\sim {\beta_f \over \beta^2}. 
\eeq

A strong suppression factor is due to the fact that the MBH has to interact with two quarks. The
cross section for accreting  a  ``di-quark first'' is suppressed by the probability for having a point-like configuration of two quarks of
size $R_{SW}$:  
\beq\label{eq-p-2}
    f_{\rm 2q} \sim (R_{\rm SW}/R_{N})^3\sim 10^{-9}.
\eeq 
(Note that this suppression is not operative for the very high energy MBHs produced by cosmic rays striking a
nucleon, considered in GM, which accrete a quark and emerge as jets.)

Some further suppression is caused by the existence of the excited colored MBH with extra energy $\alpha_s/R$ for a
time $\Delta t\sim R_{SW}/{\alpha_s^2}$. This provides the perturbative factor of 
\beq 
\label{per-fac} f_{\rm per} \sim\alpha_s \sim 0.1. 
\eeq

Collecting all factors above, that is, using Eqs. (\ref{g-cs}), (\ref{kin-sup}), (\ref{eq-p-2}), and (\ref{per-fac}) we obtain 
\beq \label{ac-cs}
  \sigma_{\rm ac} \sim \sigma_{\rm ac}^{\rm geo} \times f_{\rm kin}
\times f_{\rm 2q} \times f_{\rm per}
 \sim {10^{-43}\over \beta^2}\;{\rm cm}^2.
\eeq

Elastic scattering cross sections and resulting energy losses are dominated by Rutherford-like gravitational interactions.
The cross section is enhanced by the weight of the atom, and scales like $A^4$, where $A$ is the atomic mass. We
consider collisions between a MBH of mass $M$ and a much lighter target of mass $m_T=A m_N$. The momentum transfer is
$q \sim m_T \,\beta$ and the energy loss of the MBH is of order \beq \Delta E_{\rm sc} \sim m_N A \beta^2. \eeq

When considering the elastic cross section, for simplicity we make the conservative assumption of neglecting any
target form factors. The non-relativistic scattering amplitude can be written as \beq {\cal M} \approx {m_T\over2\pi}
\int{V d^3r}= {m_T\over2\pi}\int_R^a {d^3r}\, G_N \,M \,m_T \,{a^{d-3}\over r^{d-2}}, \eeq where we use $m_T$ as the
reduced mass, $d$ is the number of space dimensions, $a$ is the distance where the gravitational potential changes from
its $4d$ form, $V \sim r^{-1}$, to $V\sim {r^{-(d-2)}}$, which is appropriate for $d$ space dimensions. The relation
between the $4d$ and the full theory parameters is~\cite{ArkaniHamed:1998rs}
\beq 
L_{\rm Pl}^2a^{(d-3)} = (L^*_{\rm Pl})^{2+(d-3)}. 
\eeq 
This leads to 
\beq 
{\cal M} \sim A^2 m_N^2 (L^*_{\rm Pl})^3. 
\eeq 
Using $ L^*_{\rm Pl} = \TeV^{-1}$, the corresponding scattering total cross section is then 
\beq 
\label{sc-cs} \sigma_{\rm sc} \sim 4\pi|{\cal M}|^2 
\sim 4\pi A^4 {\GeV^4 \over \TeV^6}\sim 5 \times 10^{-45}\;A^4\; {\rm cm}^2. 
\eeq

\section{The Earth and the Sun}

Equipped with the relevant cross sections and energy transfers, we can check the fate of a non-physical, slow, accreting  MBH. That
is, what are its chances to escape the Earth and the solar system?

We start with the Earth. We consider an extremely slow, earth-bound MBH with $\beta \lsim 3 \times 10^{-5}$. From Eq.~(\ref{ac-cs}) 
\beq
\label{eq-sig-ac-slow}
  \sigma_{\rm ac} \sim 10^{-34}  \,{\rm cm}^2.
\eeq 
The MBH then has a mean-free path for accretion in material of density $\rho$
\beq
\label{mfp} 
L_{ac}= {1\over n \,\sigma_{\rm ac}} \approx {2 \times 10^5 \over \rho}\; {{\rm Km} \over {\rm cm^3}}, 
\eeq 
where $n = \rho \times N_{\rm Avogadro}$. Using the average earth density, $\rho \sim 5.5\;{\rm cm}^3$, we have 
\beq 
L_{ac} \sim 4 \times 10^4 \;{\rm Km} > R_E, 
\eeq 
where $R_E\approx 6.5 \times 10^3\;$Km is the Earth radius.  For the elastic scattering cross section in the
Earth we assume the Earth is pure iron, that is, we take $A=56$. Using Eq.~(\ref{sc-cs}) we get 
\beq
\label{sc-cs-earth} 
\sigma_{\rm sc} \sim 2 \times 10^{-38}\;{\rm cm}^2. 
\eeq 
We see that in the Earth the accretion
cross section, (\ref{eq-sig-ac-slow}), is much larger than the scattering cross section, (\ref{sc-cs-earth}). In particular, the
mean-free path for elastic collision is much larger than the Earth's radius. Thus we neglect the elastic collisions and
consider only accretion. Since the accretion mean-free path, $L_{\rm ac}$, is larger than the Earth's radius, most
of the slow MBHs will escape the Earth already after the first accretion. If, however, this does not happen, then in
subsequent accretions the total recoil momentum keeps ``random walking'' and on average the recoil energy builds up.  This
build-up is given by Eq.~(\ref{gen-acc}), and in the case of the Earth, the first term dominates the other two.  Thus,
the MBH keeps on gaining energy and we find that the MBH would escape the Earth after at most few accretions. This is
one of our main results.

Next we study non-physical MBHs in the Sun. MBHs ejected from Earth with $v > 45\;{\rm Km}/{\rm sec}$ escape the solar
system. Even if they escape the Earth with smaller velocity, many of them will remain bound in Keplerian orbits in the
solar system.  We have to consider, however, the small fraction, $f$, of the MBHs that can hit the sun. High
energy MBHs, whose orbits are hardly affected by the gravitational field of the Sun, are not a problem.
First, only a very few of them, $f \sim (R_{\rm Sun}/1\,{\rm A.U.})^2 \sim 5 \times 10^{-6}$, will hit the sun. More
importantly, as noted by GM, these do not stop in the sun anyway.

For those MBHs which have been produced initially (or ejected due the recoil in the first few accretions from Earth)
with small velocity, $f$ is considerably enhanced by about a factor of $(v^*_{\rm sun}/v)^2$. Here $v$ is the MBH
velocity and $v^*_{\rm sun}\sim 600\; {\rm Km}/{\rm sec}$ is the escape velocity from the surface of the Sun. This
gravitational focusing results in an enhanced fraction of slow MBHs hitting the Sun, $f \sim 10^{-3}$.

Once the MBH reach the sun its velocity is much larger due to the gravitational potential. At the surface of the Sun
the velocity is increased by about 600 Km/sec while at the center of the Sun (assuming no collision on the way) it
reaches a velocity of about 1300 Km/sec. The issue at hand is thus to find out if interaction of the MBH can change its velocity so that it will escape from the Sun.

In order to get the relevant cross sections we use $\beta \sim 3
\times 10^{-3}$.  We also recall that the Sun is mainly hydrogen with
about $1\%$ (by weight) Iron. Despite the fact that the sun has small
amount of Iron, due to the coherent enhancement the elastic cross
section is dominated by Iron. (In the Sun's core a sizable fraction of
the mass is He$^4$, but this does not change the conclusion that Iron
dominates the scattering cross section.) Using (\ref{sc-cs}) and the
fact that Iron is $1\%$ of the Sun we get the weighted average cross
section
\beq 
\sigma_{\rm sc} \sim 10^{-40}\;{\rm cm}^2. 
\eeq 
This value has to be compared with the accretion cross section in the Sun, which is mainly for protons. Using
Eq.~(\ref{ac-cs}) we obtain 
\beq 
\sigma_{\rm ac} \sim 10^{-38}\;{\rm cm}^2. 
\eeq 
Hence also in the Sun, scattering
can be neglected compared to accretions. Using Eq.~(\ref{mfp}) and $\rho\approx 1.4 g/{\rm cm}^3$ we find that the mean-free path is $10^9\,{\rm Km} \sim 10^3\, R_{\rm sun}$. Further focusing towards the center of the Sun due to the
increasing $v_{escape}$ from about 600 Km/sec at $r= R_{\rm sun}$ to about 1300 Km/sec at the center of the Sun, may
increase the effective density encountered and somewhat decrease the mean-free path.

While most of the MBHs will not interact in during their first passage
through the Sun, they will be in Keplerian orbits and keep on
returning to the sun. Eventually, after about $10^4$ years they will
interact in the Sun. The energy transfer in an accretion in the Sun is
given by (\ref{gen-acc}). Since the velocity of the MBH while in the
sun is large, the second term in (\ref{gen-acc}), that is, ${\vec
p\cdot \vec k/ M}$ is dominant. In about half of the cases the MBH
will increase its velocity and will escape the Sun and the solar
system. In the remaining cases, however, the MBH will slow down, and
will keep traversing the sun until it interacts again. We need about
$10^6$ accretions in order for the MBH to become as large as the
proton.  Then, it is more likely for the MBH to accrete the whole
proton rather then the di-quark inside it. Effectively, however, after
about $10^4$ accretions the recoil of the MBH becomes very small. Therefore, we
assume that if after the first $10^4$ interactions the MBH did not
escape from the sun, it will stay there. 

The question is, therefor, what is the probability for the MBH not to
escape the sun within its first $10^4$ interactions. The answer is
that the probability is exponentially small. Roughly, it takes about
$N^2$ steps in a random walk problem to reach a distant $N$. Thus, if
we start at the center of the sun where a recoil energy is about
$10^{-1}$ of the gravitational energy, it takes about $10^2$
accretions to escape the sun. Thus, the probability to stay after about
$10^4$ is exponentially small. To verify this argument we performed a
simple Monte Carlo simulation and found that in about $10^{-4}$ of the
cases the MBH will not escape the Sun within its first $10^3$
interactions. Our simulation is very simple, and we did not include
planetary perturbation. Such perturbations can be very important as
they can shift the orbit of these highly eccentric orbits, such that
they will miss the Sun. Performing a detailed simulation is beyond the
scope of this work. Recalling that only very small fractions of the
MBHs are produced with small velocities, and most of them will escape
the solar system in the first place, we conclude that also the Sun is
safe.

\section{Conclusions}

Micro Black Holes are new states where quantum gravity is expected to play a significant role. Their name, however,
should not mislead one to think that they are classical black holes. That is, they are quantum states and thus should
obey the laws of quantum mechanics. In particular, they must decay, and thus they cannot be problematic.

Even if one assumes that they do not decay, data show that they cannot be problematic~\cite{Giddings:2008gr}. Our work
provides one more aspect of the peaceful nature of such hypothetical MBHs. The point is that we use QCD to study how
the MBHs interact. QCD is a well proven theory and thus the assumption is well founded.

Our main observation is that the first accretions by the MBH are accompanied by a pion emission. This pion emission
results in a kick to the MBH which sends it out of the Earth and the solar system.

\section*{Acknowledgements}
We thank Aharon Casher for reminding us that the Sun also exists. The work of YG is supported by the NSF grant
PHY-0757868.



\begin{thebibliography}{99}

\bibitem{ArkaniHamed:1998rs} N.~Arkani-Hamed, S.~Dimopoulos and G.~R.~Dvali, 
Phys.\ Lett.\  B {\bf 429}, 263 (1998) [arXiv:hep-ph/9803315].

\bibitem{Antoniadis:1998ig} I.~Antoniadis, N.~Arkani-Hamed, S.~Dimopoulos and G.~R.~Dvali,
Phys.\ Lett.\  B {\bf 436}, 257 (1998)
[arXiv:hep-ph/9804398]. 

\bibitem{Randall:1999ee} L.~Randall and R.~Sundrum, 
Phys.\ Rev.\ Lett.\  {\bf 83}, 3370 (1999) [arXiv:hep-ph/9905221]. 

\bibitem{Giddings:2001bu} S.~B.~Giddings and S.~D.~Thomas, 
Phys.\ Rev.\  D {\bf 65}, 056010 (2002) [arXiv:hep-ph/0106219].

\bibitem{Dimopoulos:2001hw} S.~Dimopoulos and G.~L.~Landsberg, 
Phys.\ Rev.\ Lett.\  {\bf 87}, 161602 (2001) [arXiv:hep-ph/0106295]. 


\bibitem{Meade:2007sz}
  P.~Meade and L.~Randall,
  JHEP {\bf 0805}, 003 (2008)
  [arXiv:0708.3017 [hep-ph]].


\bibitem{Giddings:2008gr}
  S.~B.~Giddings and M.~L.~Mangano,
  Phys.\ Rev.\  D {\bf 78}, 035009 (2008)
  [arXiv:0806.3381 [hep-ph]].



\end{thebibliography}
\end{document}